\documentclass[12pt]{article}
\usepackage{amssymb}
\usepackage{amsmath}
\begin{document}

\title{Reduction and approximation in gyrokinetics}
\author{Philippe Ghendrih\thanks{%
Association Euratom-CEA, DRFC/DSM/CEA Cadarache, 13108 St. Paul lez Durance
C\'{e}dex, France; philippe.ghendrih@cea.fr}, Ricardo Lima\thanks{%
Centre de Physique Th\'{e}orique, CNRS Luminy, case 907, F-13288 Marseille
Cedex 9, France; lima@cpt.univ-mrs.fr} \hspace{0.15cm}and R. Vilela Mendes%
\thanks{%
Centro de Fus\~{a}o Nuclear - EURATOM/IST Association, Instituto Superior
T\'{e}cnico, Av. Rovisco Pais 1, 1049-001 Lisboa, Portugal} \thanks{%
CMAF, Complexo Interdisciplinar, Universidade de Lisboa, Av. Gama Pinto, 2 -
1649-003 Lisboa (Portugal), http://label2.ist.utl.pt/vilela/} \thanks{%
Corresponding author; e-mail: vilela@cii.fc.ul.pt, vilela@cpt.univ-mrs.fr}}
\maketitle
\date{}

\begin{abstract}
The gyrokinetics formulation of plasmas in strong magnetic fields aims at
the elimination of the angle associated with the Larmor rotation of charged
particles around the magnetic field lines. In a perturbative treatment or as
a time-averaging procedure, gyrokinetics is in general an approximation to
the true dynamics. Here we discuss the conditions under which gyrokinetics
is either an approximation or an exact operation in the framework of
reduction of dynamical systems with symmetry.
\end{abstract}

\section{Introduction}

In contrast with the many types of approximation schemes used to deal with
complex problems, a \textit{reduction} is an exact process. Reduction is a
process by which, given a multidimensional dynamical system, one focus on a
small subset of variables, obtaining exact equations for the variables in
this subset. Along the way, of course, one looses information about some of
the dynamical details of the system. In particular, several distinct
evolutions of the whole system may lead to the same dynamical trajectory
when projected on the space of the reduced variables but, in any case, one
is confident that the dynamics of the reduced variables is exact.

Reduction in mechanics had its origins in the classical works of Euler,
Lagrange, Hamilton, Jacobi, Routh and Poincar\'{e}. Routh's elimination of
cyclic variables and Jacobi's elimination of the node are among the first
examples (for reviews of the historical aspects and the several types of
reduction we refer to \cite{Marsden1} \cite{Landbook} \cite{Ortega}). In
general, reduction applies when the system under study possesses some kind
of symmetry. Then, the variables associated to the invariance directions are
eliminated and one ends up with a dynamical description on a quotient space.

In the context of symplectic manifolds, a construction has now become
standard \cite{Marsden2} \cite{Ortega} which, for reference and later use,
we now summarize:

Let $\left( P,\omega \right) $ be a symplectic manifold where a Lie group $G$
acts by symplectomorphisms $\left( \phi _{g}^{*}\omega =\omega \right) $. An
equivariant moment map for the group action is a map $J:P\rightarrow g^{*}$ (%
$g^{*}$ being the dual of the Lie algebra $g$ of $G$) such that if $%
\stackrel{\symbol{94}}{J}$ denotes the dual map from $g$ to the space of
smooth functions on $P$, we have 
\begin{equation}
d\left( \stackrel{\symbol{94}}{J}\left( \xi \right) \right) =i_{\xi
_{P}}\omega  \label{1.1}
\end{equation}
and $J\left( \phi _{g}\left( x\right) \right) =Ad_{g^{-1}}^{*}\left( J\left(
x\right) \right) $ , $\forall x\in P,g\in G$. If the momentum map is not
equivariant it can be converted into an equivariant one by central extension
of the group \cite{Souriau}. Eq.(\ref{1.1}) means that each infinitesimal
generator $\xi _{P}$ of $g$ has $\stackrel{\symbol{94}}{J}\left( \xi \right) 
$ as an Hamiltonian function.

Let $\mu $ be an element of $g^{*}$ and $G_{\mu }$ its coadjoint isotropy
subgroup. For any $G-$invariant Hamiltonian, $J^{-1}\left( \mu \right)
\subset P$ is an invariant set for the dynamics and the reduced space $%
J^{-1}\left( \mu \right) /G_{\mu }=P_{\mu }$ is a symplectic manifold with
symplectic form $\Omega _{\mu }$ determined by $\pi _{\mu }^{*}\Omega _{\mu
}=i_{\mu }^{*}\Omega $. $\pi _{\mu }$ is the projection $J^{-1}\left( \mu
\right) \rightarrow P_{\mu }$ and $i_{\mu }$ the inclusion $J^{-1}\left( \mu
\right) \rightarrow P$.

In the symplectic reduction the emphasis is on the projection of the Poisson
structure and Hamiltonian dynamics to a quotient space by the action of a
symmetry group. In the Lagrangian approach, reduction follows a different
approach. Reduced variables are identified and then one proceeds to carry
the variational structure to a quotient space. In the process not only the
completeness of the reduced variables set should be checked, but also
whether the variational structure can be carried over to the quotient space.
Here, in principle, a symmetry group need not to be known to begin with. One
may start by conjecturing some tentative set of reduced variables (functions
of the original variables). Then, using the Poisson brackets of the original
variables, check whether this trial set is algebraically closed. If not, the
Poisson brackets themselves will suggest new variables to close the set.
Then, to be sure that the variational structure is carried over to the
quotient space, it is necessary to check whether the (exact) dynamics of the
reduced variables may be obtained from a Lagrangian written on these
variables alone. If that is possible, an exact reduction has been performed.

Even when an exact reduction on some subset of variables is not possible,
the reduction point of view provides a step by step approach to an
approximation scheme. For example, it is possible that the trial set of
reduced variables is algebraically closed with a consistent bracket
structure, but that their dynamics cannot be entirely defined by an action
principle containing only these variables. To proceed it must be necessary
at this step to propose some kind of approximation, but at least the method
makes it clear how much of the reduction process is exact and how much it
implies an approximation.

The same methodology of approximation through reduction may be used in the
group theoretical setting. Suppose that some dynamical process is known that
is not exactly symmetric, but we are interested only on the dynamical
features consistent with the symmetry. Then an \textit{%
approximation-through-reduction} may be obtained by projecting the dynamics
on the space of functions that possess the required symmetry.

In the context of plasma physics in strong magnetic fields, two very
different time scales rule the physical phenomena. One is the fast time
scale of gyromotion of the ions around the magnetic field (of order $\frac{%
2\pi mc}{eB}$) and the other the longer time scale associated to the
electric and magnetic gradient and curvature drifts. An important
simplification arises when the two time scales can be treated separately.
This might not be possible, for example in the study of fluctuation
phenomena. However, for some stability and transport problems it is indeed
an useful approximation. When the fast time scale is separated, or
integrated over, one obtains the so called gyrokinetic and gyrocenter
equations. The usual way this is done is either by simple averaging the
single particle dynamics over the gyroperiod \cite{Catto1} \cite{Catto2} or,
more accurately, by performing a Lie-transform perturbative expansion to
obtain an action two-form where the gyroangle dependence is to be
asymptotically eliminated. In both cases the point of view is to address the
gyrokinetics formulation as an approximation to the plasma dynamics. The
Lie-transform perturbative approach has been extensively developed \cite
{Littlejohn1} \cite{Littlejohn2} \cite{Cary} \cite{Brizard1} leading, for
example, to a third-order perturbative analysis of a plasma moving with a
nonuniform fluid velocity \cite{Brizard2}.

Here, our purpose is to find out how much of the gyrokinetics program may be
framed as an exact reduction and how much it is an approximation. In Sect.2
we deal with Lagrangian reduction as a time-averaging process, following a
procedure quite analogous to the one that leads from the Lagrangian dynamics
of many particles to the Euler equation. The only difference is that now we
define time-averaged Eulerian coordinates.

The Lie-transform approach to gyrokinetics (\cite{BrizHahm} and references
therein) has been extensively studied and provides successive approximations
of practical interest to the reduced Vlasov and Maxwell equations. Its
purpose is to obtain the asymptotic elimination of the gyroangle dependence.
However to show that this procedure provides an exact reduction of the
dynamics would require the proof of convergence of the perturbative series,
what has not yet been done. In Sect.3, following a non-perturbative
approach, we show how gyrokinetics may be considered an exact reduction of
the dynamics in the sense of the symplectic reduction scheme discussed above 
\cite{Marsden2}.

\section{Lagrangian time-averaging reduction}

Here we discuss, gyrokinetics reduction as a time-averaging process, in
particular to exhibit the meaning and shortcomings of such an approach. We
start from the non-relativistic particles plus electromagnetic fields
Lagrangian \cite{Low} 
\begin{eqnarray}
L &=&\int dz_{0}f\left( z_{0}\right) \left[ \frac{m}{2}\stackrel{\bullet }{r}%
^{2}\left( z_{0},t\right) -e\int dx\left( \Phi \left( x,t\right) -\frac{%
\stackrel{\bullet }{r}\left( z_{0},t\right) }{c}\bullet A\left( x,t\right)
\right) \delta \left( x-r\left( z_{0},t\right) \right) \right]  \nonumber \\
&&+\frac{1}{8\pi }\int dx\left( \left| E\left( x,t\right) \right|
^{2}-\left| B\left( x,t\right) \right| ^{2}\right)  \label{2.1}
\end{eqnarray}
$z_{0}=\left( r_{0},p_{0}\right) $ stands for a particle coordinate in phase
space, $r\left( z_{0},t\right) $ is the position of the particle that at
time zero was at $z_{0}$, and $f\left( z_{0}\right) $ is the particle
density at time zero, that is 
\begin{equation}
f\left( z_{0}\right) =\sum_{i=1}^{N}\delta \left( r_{0}-r_{0}^{\left(
i\right) }\right) \delta \left( p_{0}-p_{0}^{\left( i\right) }\right)
\label{2.2}
\end{equation}
for $N$ particles. We consider only one ion species with mass $m$ and charge 
$e$. Generalization of all statements to an ensemble of different ion
species is straightforward.

The Lagrangian (\ref{2.1}) contains information about the identity and
dynamics of each particle at all times. We now define our reduced variables 
\begin{equation}
f_{h}\left( R,P,t\right) =\int dz_{0}f\left( z_{0}\right) \delta \left( R-%
\overline{r}\left( z_{0},t\right) \right) \delta \left( P-\frac{\overline{%
\pi }\left( z_{0},t\right) }{f\left( z_{0}\right) }\right)  \label{2.3}
\end{equation}
where 
\begin{eqnarray}
\overline{r}\left( z_{0},t\right) &=&\frac{1}{\lambda _{t}}\int d\eta
h_{t}\left( \eta \right) r\left( z_{0},t-\eta \right)  \label{2.4} \\
\overline{\pi }\left( z_{0},t\right) &=&\frac{1}{\lambda _{t}}\int d\eta
h_{t}\left( \eta \right) \pi \left( z_{0},t-\eta \right)  \nonumber
\end{eqnarray}
The kernel $h_{t}\left( \eta \right) $ is a function equal to one in the
interval $\left[ -\frac{\lambda _{t}}{2},\frac{\lambda _{t}}{2}\right] $ and
zero otherwise. The reason for the division by $f\left( z_{0}\right) $ in
the second delta function of Eq.(\ref{2.3}) comes from the fact that the
canonical momentum is 
\begin{equation}
\pi \left( z_{0},t\right) =\frac{\delta L}{\delta \stackrel{\bullet }{r}%
\left( z_{0},t\right) }=f\left( z_{0}\right) \left( m\stackrel{\bullet }{r}%
\left( z_{0},t\right) +\frac{e}{c}A\left( r\left( z_{0},t\right) \right)
\right)  \label{2.5}
\end{equation}

In the new variables two types of information are lost. First, the particle
identities are lost, being replaced by densities at points in phase space.
Second, the dynamics itself is averaged around each time $t$ with a window
of size $\lambda _{t}$, that need not be the same for all $t$.

The next step is to check the consistency of this reduction. First one
checks whether the new set $\left\{ f_{h}\left( R,P,t\right) \right\} $ is
algebraically closed and whether its Poisson bracket may be expressed on the
variables $\left( R,P\right) $. We denote $Z=\left( R,P\right) $ and compute
the Poisson bracket of the $f^{^{\prime }}s$ in the Lagrangian variables $%
\left( r,\pi \right) $. 
\begin{equation}
\begin{array}{l}
\left\{ f_{h}\left( Z,t\right) f_{h}\left( Z^{^{\prime }},t\right) \right\}
_{\left( r,\pi \right) }= \\ 
=\frac{1}{\lambda _{t}^{2}}\int dz_{0}f\left( z_{0}\right) \left[ \delta
^{^{\prime }}\left( \overline{r}\left( z_{0},t\right) -R\right) \delta
\left( \frac{\overline{\pi }}{f\left( z_{0}\right) }-P\right) \delta \left( 
\overline{r}-R^{^{\prime }}\right) \delta ^{^{\prime }}\left( \frac{%
\overline{\pi }}{f\left( z_{0}\right) }-P^{^{\prime }}\right) \right. \\ 
\hspace{3cm}\left. -\delta \left( \overline{r}\left( z_{0},t\right)
-R\right) \delta ^{^{\prime }}\left( \frac{\overline{\pi }}{f\left(
z_{0}\right) }-P\right) \delta ^{^{\prime }}\left( \overline{r}-R^{^{\prime
}}\right) \delta \left( \frac{\overline{\pi }}{f\left( z_{0}\right) }%
-P^{^{\prime }}\right) \right] \\ 
=\frac{1}{\lambda _{t}^{2}}\left[ f_{h}\left( R^{^{\prime }},P,t\right)
\delta ^{^{\prime }}\left( R^{^{\prime }}-R\right) \delta ^{^{\prime
}}\left( P-P^{^{\prime }}\right) -f_{h}\left( R,P^{^{\prime }},t\right)
\delta ^{^{\prime }}\left( R-R^{^{\prime }}\right) \delta ^{^{\prime
}}\left( P^{^{\prime }}-P\right) \right] \\ 
=\frac{1}{\lambda _{t}^{2}}\int dZ^{^{\prime \prime }}f_{h}\left(
Z^{^{\prime \prime }},t\right) \left\{ \delta \left( Z-Z^{^{\prime \prime
}}\right) ,\delta \left( Z^{^{\prime }}-Z^{^{\prime \prime }}\right)
\right\} _{\left( R,P\right) }
\end{array}
\label{2.7}
\end{equation}
One sees that not only is the set $\left\{ f_{h}\left( R,P,t\right) \right\} 
$ algebraically closed, but also its Poisson bracket may be written in the
variables $\left( R,P\right) $. The conclusion is that this set of
(Eulerian) time-averaged densities is a kinematically consistent reduction.
The quantities $R$ and $P$ are in some sense coordinates in a time-averaged
phase-space. However, one should remember that, from the reduction point of
view it is the $f^{^{\prime }}s$ that are the actual dynamical variables
which, with its bracket and reduced Hamiltonian, control the dynamics. The
set $\left( R,P\right) $ is just a set of labels for the dynamical variables.

The next step in the reduction process is to check the dynamical law and, in
particular, whether the dynamics may be completely expressed in the new
variables. The Hamiltonian that follows from (\ref{2.1}) by the Legendre
transform is 
\begin{eqnarray}
H &=&\int dz_{0}f\left( z_{0}\right) \left\{ \frac{1}{2m}\left| \frac{\pi
\left( z_{0},t\right) }{f\left( z_{0}\right) }-\frac{e}{c}A\left( r\left(
z_{0},t\right) \right) \right| ^{2}+e\Phi \left( r\left( z_{0},t\right)
\right) \right\}  \nonumber \\
&&+\frac{1}{8\pi }\int dx\left( \left| E\left( x,t\right) \right|
^{2}+\left| B\left( x,t\right) \right| ^{2}\right)  \label{2.8a}
\end{eqnarray}
Then we compute the Poisson bracket in the original Lagrangian coordinates.
Averaging the computation over the interval $\tau \in [t-\frac{\lambda _{t}}{%
2},t+\frac{\lambda _{t}}{2}]$ , considering at each $\tau $ a complete set
of canonical variables $\left( r\left( z_{0},\tau \right) ,\pi \left(
z_{0},\tau \right) \right) $. 
\begin{equation}
\begin{array}{l}
\stackrel{\bullet }{f_{h}}\left( Z,t\right) =\left\{ f_{h}\left( Z,t\right)
,H\right\} _{\left( r,\pi \right) }=\frac{1}{\lambda _{t}}\int d\tau \left\{
f_{h}\left( Z,t\right) ,H\left( \tau \right) \right\} _{\left( r\left(
z_{0},\tau \right) ,\pi \left( z_{0},\tau \right) \right) } \\ 
=\frac{1}{\lambda _{t}^{2}}\int d\tau \int dz_{0}f\left( z_{0}\right) \left[
\delta ^{^{\prime }}\left( \overline{r}\left( z_{0},t\right) -R\right)
\delta \left( \frac{\overline{\pi }\left( z_{0},t\right) }{f\left(
z_{0}\right) }-P\right) \cdot \frac{1}{m}\left( \frac{\pi \left( z_{0},\tau
\right) }{f\left( z_{0}\right) }-\frac{e}{c}A\left( z_{0},\tau \right)
\right) \right. \\ 
\hspace{1cm}+\left. \delta \left( \overline{r}\left( z_{0},t\right)
-R\right) \delta ^{^{\prime }}\left( \frac{\overline{\pi }\left(
z_{0},t\right) }{f\left( z_{0}\right) }-P\right) \cdot \left( \frac{e}{mc}%
\left( \frac{\pi \left( z_{0},\tau \right) }{f\left( z_{0}\right) }-\frac{e}{%
c}A\left( z_{0},\tau \right) \right) \cdot \nabla A-e\nabla \phi \right)
\right]
\end{array}
\label{2.9}
\end{equation}
We see that, by taking averages over $\tau $ and using the multiplication by
the delta function in the first term, one obtains the evolution equation in
the renormalized time $\frac{t}{\lambda _{t}}$%
\begin{equation}
\frac{\partial f_{h}\left( Z,t\right) }{\partial \left( \frac{t}{\lambda _{t}%
}\right) }+\left( \frac{P}{m}-\frac{e}{mc}\overline{A}\right) \cdot \frac{%
\partial f_{h}\left( Z,t\right) }{\partial R}+\overline{\left( \frac{e}{mc}%
\left( \frac{\pi }{f\left( z_{0}\right) }-\frac{e}{c}A\right) \cdot \nabla
A-e\nabla \phi \right) }\cdot \frac{\partial f_{h}\left( Z,t\right) }{%
\partial P}=0  \label{2.10}
\end{equation}
In this equation the coefficient of the $\frac{\partial }{\partial R}$
derivative takes the usual form for a Vlasov equation in time-averaged
variables but, unless the average of the products coincide with the product
of the averages, that is not the case for the coefficient of the $\frac{%
\partial }{\partial P}$ term. For the reduction to be exact one should be
able to express the dynamics of the reduced densities $f_{h}\left(
Z,t\right) $ purely in terms of the variables $\left( R,P\right) $ and the
time-averaged field potentials $\left( \overline{\phi },\overline{A}\right) $%
. It is at this step that the \textit{reduction} may become an \textit{%
approximation}. However, if it is possible, by an adequate choice of the
time-dependent averaging interval $\lambda _{t}$, to insure that the average
of the products in (\ref{2.10}) may be written in the averaged variables,
then one has an exact reduction. Otherwise the reduced dynamics implies an
approximation.

As an example of a particular situation where time-averaging provides an
exact reduction consider constant electromagnetic fields with the uniform
magnetic field $B$ oriented along the $z-$axis. The particle equations of
motion are 
\begin{equation}
\begin{array}{lll}
m\frac{dv_{z}}{dt} & = & eE_{z} \\ 
m\frac{dv_{x}}{dt} & = & eE_{x}+\frac{e}{c}v_{y}B \\ 
m\frac{dv_{y}}{dt} & = & eE_{y}-\frac{e}{c}v_{x}B
\end{array}
\label{2.11}
\end{equation}
Then 
\begin{equation}
\begin{array}{lll}
v_{x} & = & \frac{c}{B}\left( 1-\cos \left( \omega t\right) \right) E_{y}+%
\frac{1}{B}\left( cE_{x}+v_{y}\left( 0\right) B\right) \sin \left( \omega
t\right) \\ 
&  & +v_{x}\left( 0\right) \cos \left( \omega t\right) \\ 
v_{y} & = & -\frac{c}{B}\left( 1-\cos \left( \omega t\right) \right) E_{x}+%
\frac{1}{B}\left( cE_{y}-v_{x}\left( 0\right) B\right) \sin \left( \omega
t\right) \\ 
&  & +v_{y}\left( 0\right) \cos \left( \omega t\right) \\ 
v_{z} & = & \frac{e}{m}tE_{z}+v_{z}\left( 0\right)
\end{array}
\label{2.12}
\end{equation}
$\omega =\frac{e\left| B\right| }{mc}$. Choosing $\lambda =\frac{2\pi }{%
\omega }$, the corresponding averaged dynamics is 
\begin{equation}
\begin{array}{cll}
\overline{v}_{x} & = & c\frac{\overline{E}_{y}}{B} \\ 
\overline{v}_{y} & = & -c\frac{\overline{E}_{x}}{B} \\ 
\overline{v}_{z} & = & \frac{e}{m}t\overline{E}_{z}+\overline{v}_{z}\left(
0\right) \\ 
\frac{d\overline{v}_{x}}{dt} & = & 0 \\ 
\frac{d\overline{v}_{y}}{dt} & = & 0 \\ 
\frac{d\overline{v}_{z}}{dt} & = & \frac{e}{m}\overline{E}_{z}
\end{array}
\label{2.13}
\end{equation}
and the reduced Vlasov equation 
\begin{equation}
\frac{\partial f}{\partial t}+\overline{v}_{x}\frac{\partial f}{\partial 
\overline{x}}+\overline{v}_{y}\frac{\partial f}{\partial \overline{y}}+%
\overline{v}_{z}\frac{\partial f}{\partial \overline{z}}+\frac{e}{m}%
\overline{E}_{z}\left( f\right) \frac{\partial f}{\partial \overline{v}_{z}}%
=0  \label{2.14}
\end{equation}
As expected in an exact reduction, the angle, argument of the trigonometric
functions, being an ignorable coordinate for the averaged dynamics, the
number of relevant coordinates is reduced from six, $\left(
x,y,z,v_{x},v_{y},v_{z}\right) $, to four, $\left( \overline{x},\overline{y},%
\overline{z},\overline{v}_{z}\right) $.

Here we have analyzed reduction as a time-averaging process to emphasize its
different aspects, namely the structural (or kinematical) level that
concerns the Poisson brackets and the dynamical level that refers to the
equations of motion. Although we have considered an averaging interval $%
\lambda _{t}$ that depends on time, it must be pointed out that, for general
magnetic field configurations, the gyrokinetics approximation is more
general in the sense that the averaging depends not only in time but also on
the position of the particles. In the present Lagrangian reduction scheme it
would correspond to having a kernel $h=h\left( z_{0},t\right) $.

For general electromagnetic fields, the difficulties with the time-averaging
point of view lead naturally to the more appropriate notion of gyroangle
independence, which is also the point of view of the Lie-transform
perturbative approach \cite{BrizHahm}. Here we also follow the gyroangle
independence point of view but, instead of a perturbative approach, we prove
the existence of an exact invariant from which an exact reduction is shown
to follow.

\section{Gyrokinetics as an exact reduction}

In the extended phase-space $\left( \stackrel{\rightarrow }{x},\stackrel{%
\rightarrow }{p},t,h\right) $, the Hamiltonian of a particle moving in an
electromagnetic field is \cite{Littlejohn3} 
\begin{equation}
H\left( \stackrel{\rightarrow }{x},\stackrel{\rightarrow }{p},t,h\right) =%
\frac{1}{2m}\left( \stackrel{\rightarrow }{p}-\frac{e}{c}\stackrel{%
\rightarrow }{A}\right) ^{2}+e\Phi -h  \label{3.01}
\end{equation}
the non-vanishing elements of the Poisson tensor being 
\begin{equation}
\begin{array}{lll}
\left\{ x^{i},p^{j}\right\} & = & \delta ^{ij} \\ 
\left\{ t,h\right\} & = & -1
\end{array}
\label{3.02}
\end{equation}
Changing coordinates to $\left( \stackrel{\rightarrow }{x},\stackrel{%
\rightarrow }{v},t,k\right) $ with 
\begin{equation}
\begin{array}{lll}
\stackrel{\rightarrow }{v} & = & \frac{1}{m}\left( \stackrel{\rightarrow }{p}%
-\frac{e}{c}\stackrel{\rightarrow }{A}\right) \\ 
k & = & h-e\Phi
\end{array}
\label{3.03}
\end{equation}
leads to 
\begin{equation}
H\left( \stackrel{\rightarrow }{x},\stackrel{\rightarrow }{v},t,k\right) =%
\frac{1}{2}m\stackrel{\rightarrow }{v}^{2}-k  \label{3.04}
\end{equation}
and 
\begin{equation}
\begin{array}{rll}
\sigma ^{iv^{j}}=\left\{ x^{i},v^{j}\right\} & = & \frac{1}{m}\delta ^{ij}
\\ 
\sigma ^{tk}=\left\{ t,k\right\} & = & -1 \\ 
\sigma ^{v^{i}v^{j}}=\left\{ v^{i},v^{j}\right\} & = & \frac{e}{m^{2}c}B^{ij}
\\ 
\sigma ^{v^{i}k}=\left\{ v^{i},k\right\} & = & -\frac{e}{m}E^{i}
\end{array}
\label{3.05}
\end{equation}
$B^{ij}=\epsilon ^{ijk}B_{k}$

It is convenient to decompose the velocity into magnetic field adapted
components, 
\begin{equation}
\begin{array}{lll}
v_{\Vert } & = & v_{i}\stackrel{\symbol{94}}{b_{i}} \\ 
\stackrel{\rightarrow }{v}_{\bot } & = & \stackrel{\rightarrow }{v}-%
\stackrel{\symbol{94}}{b}\left( \stackrel{\rightarrow }{v}\cdot \stackrel{%
\symbol{94}}{b}\right)
\end{array}
\label{3.06}
\end{equation}
$\stackrel{\symbol{94}}{b}=\frac{\stackrel{\rightarrow }{B}}{\left| 
\stackrel{\rightarrow }{B}\right| }$ , for which the equations of motion are
obtained from (\ref{3.04}) and (\ref{3.05}) by $\frac{dF}{dt}=\left\{
F,H\right\} $%
\begin{equation}
\begin{array}{lll}
\frac{d}{dt}v_{\Vert } & = & \frac{e}{m}E_{\Vert }+\stackrel{\rightarrow }{%
v_{\bot }}\cdot \left( \stackrel{\rightarrow }{v}\cdot \nabla \right) 
\stackrel{\symbol{94}}{b} \\ 
\frac{d}{dt}\stackrel{\rightarrow }{v}_{\bot } & = & \frac{e}{m}\stackrel{%
\rightarrow }{E}_{\bot }+\frac{e}{mc}\left( \stackrel{\rightarrow }{v}_{\bot
}\times \stackrel{\rightarrow }{B}\right) -v^{i}\left( \stackrel{\rightarrow 
}{v}\cdot \nabla \right) \left( \stackrel{\symbol{94}}{b}^{i}\stackrel{%
\symbol{94}}{b}\right)
\end{array}
\label{3.09}
\end{equation}
For bounded electromagnetic fields with bounded derivatives, the right-hand
side of the system of equations (\ref{3.09}) is locally lipschitzian. This
insures existence and uniqueness of the solution for a time interval.

To prove the existence of an exact reduction of a dynamical system one has
to identify a symmetry group or, equivalently, the existence of one or more
invariants. Our method depends on the construction of a formal invariant.
Existence of such invariants, to all orders in perturbation theory, for
charged particle motion in a strong magnetic field was first pointed out by
Kruskal\cite{Kruskal}. Here we attempt an explicit construction of an exact
invariant. The technique hinges on transforming the equation for the
transverse velocity to the form\textit{\ } 
\begin{equation}
\frac{d}{dt}\left( \stackrel{\rightarrow }{v}_{\bot }-\stackrel{\rightarrow 
}{u}_{\bot }\right) =k\left( \stackrel{\rightarrow }{v}_{\bot }-\stackrel{%
\rightarrow }{u}_{\bot }\right) \times \stackrel{\rightarrow }{B}-\stackrel{%
\symbol{94}}{b}\Gamma \left( \stackrel{\rightarrow }{x},\stackrel{%
\rightarrow }{v},t\right) -\alpha \left( \stackrel{\rightarrow }{x},%
\stackrel{\rightarrow }{v},t\right) \left( \stackrel{\rightarrow }{v}_{\bot
}-\stackrel{\rightarrow }{u}_{\bot }\right)  \label{3.011}
\end{equation}
When this is achieved it is easy to see that 
\begin{equation}
M=\frac{\left| \stackrel{\rightarrow }{v}_{\bot }-\stackrel{\rightarrow }{u}%
_{\bot }\right| ^{2}}{F}  \label{3.012}
\end{equation}
is a constant of motion provided 
\begin{equation}
F=\exp \left( -2\int^{t}\alpha \left( \stackrel{\rightarrow }{x},\stackrel{%
\rightarrow }{v},\tau \right) d\tau \right)  \label{3.013}
\end{equation}
The results are summarized in the following proposition:

\textit{In the domain of existence of bounded solutions of the system (\ref
{3.09}) and for bounded and sufficiently smooth electromagnetic fields,
there are functions }$\Gamma \left( \stackrel{\rightarrow }{x},\stackrel{%
\rightarrow }{v},t\right) $\textit{, }$\alpha \left( \stackrel{\rightarrow }{%
x},\stackrel{\rightarrow }{v},t\right) $\textit{\ and a transversal vector
function }$\stackrel{\rightarrow }{u}_{\bot }\left( \stackrel{\rightarrow }{x%
},\stackrel{\rightarrow }{v},t\right) $\textit{\ such that the equation (\ref
{3.011}) holds. For fields that have no explicit time dependence and
sufficiently large magnetic field we also have }$\stackrel{\rightarrow }{u}%
_{\bot }=\stackrel{\rightarrow }{u}_{\bot }\left( \stackrel{\rightarrow }{x},%
\stackrel{\rightarrow }{v}\right) $, $\alpha =\alpha \left( \stackrel{%
\rightarrow }{x},\stackrel{\rightarrow }{v}\right) $ \textit{\ and }$\Gamma
=\Gamma \left( \stackrel{\rightarrow }{x},\stackrel{\rightarrow }{v}\right) $%
\textit{.}

Proof : Without explicit time-independence of $\stackrel{\rightarrow }{u}%
_{\bot }$, $\alpha $ and $\Gamma $, the result is fairly trivial. It
suffices to rewrite (\ref{3.011}) as 
\begin{equation}
\frac{d}{dt}\stackrel{\rightarrow }{u}_{\bot }-k\stackrel{\rightarrow }{u}%
_{\bot }\times \stackrel{\rightarrow }{B}+\alpha \left( \stackrel{%
\rightarrow }{x},\stackrel{\rightarrow }{v},t\right) \stackrel{\rightarrow }{%
u}_{\bot }=\left( \frac{d}{dt}+\alpha \left( \stackrel{\rightarrow }{x},%
\stackrel{\rightarrow }{v},t\right) -k\stackrel{\rightarrow }{B}\times
\right) \stackrel{\rightarrow }{v}_{\bot }+\stackrel{\symbol{94}}{b}\Gamma
\left( \stackrel{\rightarrow }{x},\stackrel{\rightarrow }{v},t\right)
\label{3.111}
\end{equation}
and then, the assumed boundedness of the fields and the solutions of (\ref
{3.09}) implies the existence of a solution $\stackrel{\rightarrow }{u}%
_{\bot }\left( \stackrel{\rightarrow }{x},\stackrel{\rightarrow }{v}%
,t\right) $ for each choice of $\Gamma \left( \stackrel{\rightarrow }{x},%
\stackrel{\rightarrow }{v},t\right) $ and $\alpha \left( \stackrel{%
\rightarrow }{x},\stackrel{\rightarrow }{v},t\right) $.

Less trivial is to show the existence of time-independent solutions. Because
we are going to construct the quantities $\stackrel{\rightarrow }{u}_{\bot }$%
, $\alpha $ and $\Gamma $ and the invariant as a formal operator series, it
is important to control the magnitude of the operator action in the space of
velocities, in particular to guarantee that it does not grow with $\left|
B\right| $, the magnetic field intensity. In the second equation in (\ref
{3.09}) we bring the term $\frac{e}{mc}\left( \stackrel{\rightarrow }{v}%
_{\bot }\times \stackrel{\rightarrow }{B}\right) $ to the left-hand side 
\begin{equation}
\begin{array}{l}
\left( \frac{d}{dt}+\frac{e}{mc}\stackrel{\rightarrow }{B}\times \right) 
\stackrel{\rightarrow }{v}_{\bot }=\frac{e}{m}\stackrel{\rightarrow }{E}%
_{\bot }-v^{i}\left( \stackrel{\rightarrow }{v}\cdot \nabla \right) \left( 
\stackrel{\symbol{94}}{b}^{i}\stackrel{\symbol{94}}{b}\right) 
\end{array}
\label{3.09a}
\end{equation}
and rewrite the operator $\left( \frac{d}{dt}+\frac{e}{mc}\stackrel{%
\rightarrow }{B}\times \right) $ as a derivation 
\begin{equation}
D=\frac{d}{dt}+\frac{e\left| B\right| }{mc}\left( v_{\bot }^{\left( 1\right)
}\frac{\partial }{\partial v_{\bot }^{\left( 2\right) }}-v_{\bot }^{\left(
2\right) }\frac{\partial }{\partial v_{\bot }^{\left( 1\right) }}\right) 
\label{3.09b}
\end{equation}
$v_{\bot }^{\left( 1\right) }$ and $v_{\bot }^{\left( 1\right) }$ being the
components of $\stackrel{\rightarrow }{v_{\bot }}$ in an arbitrary
transverse coordinate system. Then the equations (\ref{3.09}) become 
\begin{equation}
\begin{array}{lll}
Dv_{\Vert } & = & \frac{e}{m}E_{\Vert }+\stackrel{\rightarrow }{v_{\bot }}%
\cdot \left( \stackrel{\rightarrow }{v}\cdot \nabla \right) \stackrel{%
\symbol{94}}{b} \\ 
D\stackrel{\rightarrow }{v}_{\bot } & = & \frac{e}{m}\stackrel{\rightarrow }{%
E}_{\bot }-v^{i}\left( \stackrel{\rightarrow }{v}\cdot \nabla \right) \left( 
\stackrel{\symbol{94}}{b}^{i}\stackrel{\symbol{94}}{b}\right) 
\end{array}
\label{3.09c}
\end{equation}
We see that the right-hand side no longer involves explicit dependence on
the ,magnetic field, that is, the action of the $D$ operator does not
introduce frequencies comparable to the Larmor frequency. It is the choice
of the $D$ operator for the operator expansions that, for large $B$,
implements the separation of time scales.

We now apply the identity, 
\begin{equation}
\stackrel{\rightarrow }{a}_{\perp }=-\frac{1}{\left| \stackrel{\rightarrow }{%
B}\right| ^{2}}\left( \stackrel{\rightarrow }{a}_{\perp }\times \stackrel{%
\rightarrow }{B}\right) \times \stackrel{\rightarrow }{B}  \label{5.1f}
\end{equation}
that holds for transversal fields, to the first and the transversal part of
the second term in the right-hand side of the second equation in (\ref{3.09c}%
), to obtain

\begin{equation}
\begin{array}{r}
D\left( \stackrel{\rightarrow }{v}_{\bot }-c\frac{\stackrel{\rightarrow }{E}%
_{\bot }\times \stackrel{\rightarrow }{B}}{\left| B\right| ^{2}}+\frac{%
mcv_{\shortparallel }}{e\left| B\right| ^{2}}\left( \stackrel{\rightarrow }{v%
}\cdot \nabla \right) \stackrel{\symbol{94}}{b}\times \stackrel{\rightarrow 
}{B}\right) =\frac{e}{mc}\left( -c\frac{\stackrel{\rightarrow }{E}_{\bot
}\times \stackrel{\rightarrow }{B}}{\left| B\right| ^{2}}+\frac{%
mcv_{\shortparallel }}{e\left| B\right| ^{2}}\left( \stackrel{\rightarrow }{v%
}\cdot \nabla \right) \stackrel{\symbol{94}}{b}\times \stackrel{\rightarrow 
}{B}\right) \times \stackrel{\rightarrow }{B} \\ 
-D\left\{ c\frac{\stackrel{\rightarrow }{E}_{\bot }\times \stackrel{%
\rightarrow }{B}}{\left| B\right| ^{2}}-\frac{mcv_{\shortparallel }}{e\left|
B\right| ^{2}}\left( \stackrel{\rightarrow }{v}\cdot \nabla \right) 
\stackrel{\symbol{94}}{b}\times \stackrel{\rightarrow }{B}\right\} -%
\stackrel{\symbol{94}}{b}v^{i}\left( \stackrel{\rightarrow }{v}\cdot \nabla
\right) \stackrel{\symbol{94}}{b}^{i}
\end{array}
\label{5.2}
\end{equation}
Separating the transversal and longitudinal components of the term $D\left\{
\cdots \right\} $ in the right-hand side of Eq.(\ref{5.2}), using the
identity (\ref{5.1f}) for the transversal component and iterating the
process one finally obtains 
\begin{equation}
\frac{d}{dt}\left( \stackrel{\rightarrow }{v}_{\bot }-\stackrel{\rightarrow 
}{u}_{\bot }\right) =\left( \frac{e}{mc}+\gamma \right) \left( \stackrel{%
\rightarrow }{v}_{\bot }-\stackrel{\rightarrow }{u}_{\bot }\right) \times 
\stackrel{\rightarrow }{B}-\stackrel{\symbol{94}}{b}\Gamma -\alpha \left( 
\stackrel{\rightarrow }{v}_{\bot }-\stackrel{\rightarrow }{u}_{\bot }\right)
\label{5.2a}
\end{equation}
with $\stackrel{\rightarrow }{u}_{\bot }$, $\Gamma $ and $\alpha $ given by 
\begin{equation}
\begin{array}{lll}
\stackrel{\rightarrow }{u}_{\bot } & = & \left\{ 1-\frac{mc}{e\left|
B\right| ^{2}}\stackrel{\rightarrow }{B}\times D\right\} ^{-1}\left( c\frac{%
\stackrel{\rightarrow }{E}_{\bot }\times \stackrel{\rightarrow }{B}}{\left|
B\right| ^{2}}-\frac{mcv_{\shortparallel }}{e\left| B\right| ^{2}}\left( 
\stackrel{\rightarrow }{v}\cdot \nabla \right) \stackrel{\symbol{94}}{b}%
\times \stackrel{\rightarrow }{B}\right)
\end{array}
\label{5.2b}
\end{equation}
and 
\begin{equation}
\begin{array}{rrr}
\Gamma & = & \stackrel{\symbol{94}}{b}\cdot D\left( 1-\frac{mc}{e\left|
B\right| ^{2}}\stackrel{\rightarrow }{B}\times D\right) ^{-1}\left( c\frac{%
\stackrel{\rightarrow }{E}_{\bot }\times \stackrel{\rightarrow }{B}}{\left|
B\right| ^{2}}-\frac{mcv_{\shortparallel }}{e\left| B\right| ^{2}}\left( 
\stackrel{\rightarrow }{v}\cdot \nabla \right) \stackrel{\symbol{94}}{b}%
\times \stackrel{\rightarrow }{B}\right) \\ 
&  & +\stackrel{\symbol{94}}{b}\cdot \chi +v^{i}\left( v\cdot \nabla \right) 
\stackrel{\symbol{94}}{b}^{i}
\end{array}
\label{5.2d}
\end{equation}
\begin{equation}
\alpha =\frac{\chi \cdot \left( \stackrel{\rightarrow }{v}_{\bot }-\stackrel{%
\rightarrow }{u}_{\bot }\right) }{\left| \stackrel{\rightarrow }{v}_{\bot }-%
\stackrel{\rightarrow }{u}_{\bot }\right| ^{2}}  \label{5.2d1}
\end{equation}
and 
\begin{equation}
\chi =\left( \frac{d}{dt}-D\right) \stackrel{\rightarrow }{u}_{\bot }
\label{5.2d2}
\end{equation}
\begin{equation}
\gamma =-\left( \frac{\left( \stackrel{\rightarrow }{v}_{\bot }-\stackrel{%
\rightarrow }{u}_{\bot }\right) }{\left| \stackrel{\rightarrow }{v}_{\bot }-%
\stackrel{\rightarrow }{u}_{\bot }\right| ^{2}}\times \frac{\stackrel{%
\rightarrow }{B}}{\left| B\right| ^{2}}\right) \cdot \chi  \label{5.2d3}
\end{equation}
Notice that $\frac{d}{dt}=\frac{\partial }{\partial t}+\left( \stackrel{%
\rightarrow }{v}\cdot \nabla \right) $. When $D$ is applied to the
velocities it is understood that it is a replacement by the right-hand side
of Eqs.(\ref{3.09c}). When $D$ is applied to the fields, explicit time
dependence in $\stackrel{\rightarrow }{u}_{\bot }$ and $\Gamma $ occurs only
if the fields themselves have an explicit time dependence. Therefore,
whenever a proper meaning is given to the series in (\ref{5.2b}) and the
fields are not explicitly dependent on time, one proves the existence of
time-independent solutions $\stackrel{\rightarrow }{u}_{\bot }=\stackrel{%
\rightarrow }{u}_{\bot }\left( \stackrel{\rightarrow }{x},\stackrel{%
\rightarrow }{v}\right) $,\textit{\ }$\Gamma =\Gamma \left( \stackrel{%
\rightarrow }{x},\stackrel{\rightarrow }{v}\right) $ and $\alpha \left( 
\stackrel{\rightarrow }{x},\stackrel{\rightarrow }{v}\right) $.

We now discuss the convergence of the formal series in (\ref{5.2b}). The
operator $\stackrel{\rightarrow }{B}\times D$ having the real line as
spectrum, $1-\frac{mc}{e\left| B\right| ^{2}}\stackrel{\rightarrow }{B}%
\times D$ cannot have an inverse defined in the whole $L^{2}$. Instead we
consider the action of the operator at sucessively higher orders on a space
of velocities and fields bounded by some quantity $M$. Because of the choice
of the operator $D$, the terms do not grow with $B$, however, because of the
nonlinear nature of the (\ref{3.09c}) action, there is a proliferation of
terms and, at most, we obtain a bound 
\[
\left| \stackrel{\rightarrow }{u}_{\bot }\right| \leq \sum_{n}n!\left( \frac{%
M}{\left| B\right| }\right) ^{n} 
\]
which does not insure convergence. However, multiplying and dividing by $n!$
and exchanging the order of sum and integral one obtains 
\[
\left| \stackrel{\rightarrow }{u}_{\bot }\right| \leq \int_{d}\sum_{n}\left( 
\frac{Mz}{\left| B\right| }\right) ^{n}e^{-z}dz 
\]
where we have used the integral representation 
\[
n!=\int_{d}dzz^{n}e^{-z} 
\]
$d$ denoting the integration along the line $\left( 1+\varepsilon \right)
\eta $ ($\eta $ real $\in [0,\infty )$) in the complex plane. Then 
\[
\left| \stackrel{\rightarrow }{u}_{\bot }\right| \leq
C\int_{d}\sum_{n}\left( 1-\frac{Mz}{\left| B\right| }\right) ^{-1}e^{-z}dz 
\]
meaning that $\left| \stackrel{\rightarrow }{u}_{\bot }\right| $ is bounded
by a series that is Borel summable along a direction not containing the real
half-line. Then the formal series in (\ref{5.2b}) is also expected to be
Borel summable. In this case, by a theorem of Borel \cite{Borel} \cite{Ramis}%
, one solution $ \stackrel{\rightarrow }{\underset{\symbol{126}}{u}}%
_{\bot } $ exists for which (\ref{5.2b}) is an asymptotic series. The
nature of (\ref{5.2b}) as an asymptotic series has the implication that for
practical calculations the number of terms to be kept depends on both the
magnitudes of the velocities and the magnetic field.

Now from the existence of a vector function $\stackrel{\rightarrow }{u}%
_{\bot }$, satisfying (\ref{3.011}), it follows that for $M$, constructed in
(\ref{3.012}), $\frac{d}{dt}M=0$. Therefore, an exact reduction is now
possible following the Marsden-Weinstein theory \cite{Marsden2}. $M$ itself
is the (dual) moment map (the $\stackrel{\symbol{94}}{J}\left( \xi \right) $
in Eq.\ref{1.1}) that by 
\begin{equation}
\frac{dQ}{dt}=\left\{ Q,M\right\}  \label{3.012a}
\end{equation}
generates the action of the symmetry group on phase-space functions $Q\left( 
\stackrel{\rightarrow }{x},\stackrel{\rightarrow }{v}\right) $.

For each value $\mu $ of the invariant $M$ one obtains a symplectic reduced
space of dimension four. Existence of a set $\left( \stackrel{\rightarrow }{Q%
},Q_{v}\right) $ of four $Q-$coordinates in the reduced space follows from
the existence of solutions for the linear (in $Q$) equation 
\begin{equation}
\left\{ Q,M\right\} =0  \label{3.015}
\end{equation}
that is, 
\begin{equation}
\frac{\partial M}{\partial v^{i}}\sigma ^{v^{i}j}\frac{\partial Q}{\partial
x^{j}}+\frac{\partial M}{\partial x^{i}}\sigma ^{iv^{j}}\frac{\partial Q}{%
\partial v^{j}}+\frac{\partial M}{\partial v^{i}}\sigma ^{v^{i}v^{j}}\frac{%
\partial Q}{\partial v^{j}}=0  \label{3.015b}
\end{equation}
for fields without explicit time-dependence.

For constant uniform fields one would obtain as solutions of (\ref{3.015b})
the following set of coordinates for the reduced space 
\begin{equation}
\begin{array}{lll}
\stackrel{\rightarrow }{Q}^{(0)} & = & \stackrel{\rightarrow }{x}+\frac{mc}{%
e\left| \stackrel{\rightarrow }{B}\right| ^{2}}\left( \stackrel{\rightarrow 
}{v_{\bot }}-\stackrel{\rightarrow }{u}_{\bot }\right) \times \stackrel{%
\rightarrow }{B} \\ 
Q_{v}^{(0)} & = & \stackrel{\symbol{94}}{b}\cdot \stackrel{\rightarrow }{v}
\end{array}
\label{3.017}
\end{equation}
Corrections to these coordinates for the general case arise from the
deviation from uniformity in the fields. Therefore an iteration scheme may
be devised to construct these corrections with convergence dependent on the
fast approach to zero of the higher space derivatives of the fields. Write (%
\ref{3.015b}) as 
\begin{equation}
\gamma _{i}\frac{\partial }{\partial z_{i}}Q=0  \label{5.3}
\end{equation}
with $\gamma =\left( \stackrel{\rightarrow }{\gamma _{x}},\stackrel{%
\rightarrow }{\gamma _{v}}\right) $ and $z=\left( \stackrel{\rightarrow }{x},%
\stackrel{\rightarrow }{v}\right) $

Consider now an arbitrary unit vector $\stackrel{\symbol{94}}{a}$ transverse
to $\stackrel{\symbol{94}}{b}$. We may take $\stackrel{\symbol{94}}{a}=\frac{%
c\times \stackrel{\symbol{94}}{b}}{\left| c\times \stackrel{\symbol{94}}{b}%
\right| }$, $c$ being a fixed vector in space, non collinear with $\stackrel{%
\symbol{94}}{b}$. Then define 
\begin{equation}
\theta _{0}=\frac{mc}{e\left| B\right| }\cos ^{-1}\left( \frac{\stackrel{%
\symbol{94}}{a}\cdot \left( \stackrel{\rightarrow }{v}_{\bot }-\stackrel{%
\rightarrow }{u}_{\bot }\right) }{\left| \stackrel{\rightarrow }{v}_{\bot }-%
\stackrel{\rightarrow }{u}_{\bot }\right| }\right)  \label{5.5}
\end{equation}
$\theta _{0}$ is the angle variable conjugate to $M$ in the case of uniform
fields. Indeed, in this case $\gamma _{i}\frac{\partial }{\partial z_{i}}%
\theta _{0}$ reduces to 
\begin{equation}
\left( \stackrel{\rightarrow }{v}_{\bot }-\stackrel{\rightarrow }{u}_{\bot
}\right) _{j}\left\{ \frac{\partial }{\partial x_{j}}+\frac{e}{mc}B_{kj}%
\frac{\partial }{\partial v_{k}}\right\} \theta _{0}=1  \label{5.6}
\end{equation}

We now write (\ref{5.3}) as 
\begin{equation}
\gamma _{i}\frac{\partial }{\partial z_{i}}\left( Q^{\left( 0\right)
}+Q^{\left( 1\right) }+\cdots \right) =0  \label{5.7}
\end{equation}
or 
\begin{equation}
\gamma _{i}\frac{\partial }{\partial z_{i}}\left( Q^{\left( 1\right)
}+\cdots \right) =-\gamma _{i}\frac{\partial }{\partial z_{i}}Q^{\left(
0\right) }  \label{5.8}
\end{equation}
Putting 
\begin{equation}
Q^{\left( 1\right) }=-\theta _{0}\gamma _{i}\frac{\partial }{\partial z_{i}}%
Q^{\left( 0\right) }  \label{5.9}
\end{equation}
one cancels the lowest order (in the field derivatives) terms in the
right-hand side of (\ref{5.8}). Iterating the procedure one finally obtains 
\begin{equation}
Q=\frac{1}{1+\theta _{0}\gamma _{i}\frac{\partial }{\partial z_{i}}}%
Q^{\left( 0\right) }  \label{5.10}
\end{equation}
As stated, convergence of the formal series (\ref{5.10}) will rely on the
fast convergence to zero of higher order space derivatives of the fields.
These reduced space coordinates are the variables that should enter into the
gyrokinetics reduced Vlasov equation.

In conclusion: By proving the existence of a vector function $\stackrel{%
\rightarrow }{u}_{\bot }$ satisfying Eq.(\ref{3.011}) we have somehow
reversed the usual approach to gyrokinetics. Instead of the (still open)
question of asymptotic gyroangle independence in the perturbative approach,
we have shown the existence of an exact reduction and, whenever
approximations are needed, they will focus on truncations of the exact $%
\stackrel{\rightarrow }{u}_{\bot }$ and the corresponding approximations for
the set of variables $\left( \stackrel{\rightarrow }{Q},Q_{v}\right) $ in
the reduced space.

{\large Acknowledgments}

The authors are grateful to Prof. Alain Brizard for his comments on an
earlier version of this paper.


\begin{thebibliography}{99}
\bibitem{Marsden1}  H. Cendra, J. E. Marsden and T. S. Ratiu; in \textit{%
Mathematics Unlimited-2001 and Beyond, }B. Engquist and W. Schmid (Eds.),
pp. 221-273, Springer, NY 2001.

\bibitem{Landbook}  N. P. Landsman; \textit{Mathematical topics between
classical and quantum mechanics}, Springer, New York 1998.

\bibitem{Ortega}  J.-P. Ortega and T. S. Ratiu; \textit{Momentum maps and
Hamiltonian reduction}, Progr. in Math. vol. 222, Birkh\"{a}user, Boston
2004.

\bibitem{Marsden2}  J. E. Marsden and A. Weinstein; Rep. Math. Phys. 5
(1974) 121-130.

\bibitem{Souriau}  J. M. Souriau; \textit{Structure des Syst\`{e}mes
Dynamiques}, Dunod, Paris 1970.

\bibitem{Catto1}  P. J. Catto; Plasma Physics 20 (1978) 719-722.

\bibitem{Catto2}  P. J. Catto, W. M. Tang and D. E. Baldwin; Plasma Physics
23 (1981) 639-650.

\bibitem{Littlejohn1}  R. G. Littlejohn; J. Math. Phys. 23 (1982) 742

\bibitem{Littlejohn2}  R. G. Littlejohn; J. Plasma Phys. 29 (1983) 111

\bibitem{Cary}  J. R. Cary and R. G. Littlejohn; Ann. Phys. (N. Y.) 151
(1983) 1

\bibitem{Brizard1}  A. J. Brizard; J. Plasma Phys. 41 (1989) 541

\bibitem{Brizard2}  A. J. Brizard; Phys. Plasmas 2 (1995) 459-471.

\bibitem{BrizHahm}  A. J. Brizard and T. S. Hahm; Priceton report PPPL-4153.

\bibitem{Low}  F. E. Low; Proc. R. Soc. London, Ser. A 248 (1958) 282.

\bibitem{Littlejohn3}  R. G. Littlejohn; Phys. Fluids 24 (1981) 1730-1749.

\bibitem{Marsden3}  J. E. Marsden and A. Weinstein; Physica 4D (1982)
394-406.

\bibitem{Kruskal}  M. Kruskal; J. Math. Phys. 3 (1962) 806-828.

\bibitem{Borel}  E. Borel; \textit{Le\c{c}ons sur les s\'{e}ries divergentes}%
, Gauthier-Villars 1901.

\bibitem{Ramis}  J.-P. Ramis; \textit{S\'{e}ries divergentes et th\'{e}ories
asymptotiques}, Panoramas et Synth\`{e}ses, Soc. Math. France 1994.
\end{thebibliography}
\end{document}